\def\rect#1#2{{\vcenter{\vbox{\hrule height.3pt
	    \hbox{\vrule width.3pt height#2truecm \kern#1truecm
	    \vrule width.3pt}
	    \hrule height.3pt}}}}

\def\aversim#1#2{\lower3pt\vbox{\baselineskip0pt \lineskip-.1pt
    \ialign{$\mathsurround=0pt #1\hfil##\hfil$\crcr#2\crcr\sim\crcr}}}

\documentstyle[12pt]{article}
\begin{document}
\begin{center}
\LARGE
Stochastic Learning in a Neural Network with Adapting Synapses.
~\\
~\\
\vskip 1.5 truecm
\normalsize
G. Lattanzi$^1$, G. Nardulli$^1$, G. Pasquariello$^2$ and S. Stramaglia$^2$\\
\vskip .5 cm
$^1${\it Dipartimento di Fisica dell'Universit\`{a} di Bari} and\\
{\it Istituto Nazionale di Fisica Nucleare, Sezione di Bari\\
via Amendola 173, 70126 Bari, Italy}\\
E\_mail: lattanzi@axpba4.ba.infn.it, nardulli@axpba0.ba.infn.it\\
\vskip .35 cm
$^2${\it Istituto Elaborazione Segnali ed Immagini}\\
{\it Consiglio Nazionale delle Ricerche}\\
{\it via Amendola 166/5, 70126 Bari, Italy}\\
E\_mail: guido@iesi.ba.cnr.it, sebino@iesi.ba.cnr.it\\
~\\
~\\
\end{center}
\vskip 1. cm
\begin{abstract}
We consider a neural network with adapting synapses whose dynamics can be 
analitically computed.
The model is made of $N$ neurons and each of them is connected to $K$ input 
neurons chosen at random in the network.
The synapses are  $n$-states variables  
which evolve in time according to  
Stochastic Learning rules; a parallel stochastic dynamics is assumed for 
neurons.
Since the network maintains the same dynamics whether it is engaged in 
computation or in learning new memories, a very low probability of synaptic 
transitions is assumed.
In the limit $N\to\infty$ with $K$ large and finite, 
the correlations of neurons 
and synapses can be neglected and  the dynamics
can be analitically calculated by flow equations for the macroscopic 
parameters of the system. 
\end{abstract}

\newpage
\addtolength{\baselineskip}{\baselineskip}
\par\noindent
{\bf 1. Introduction.}
\par
In the dynamics of Attractor Neural Networks (ANNs) two processes take place:
the time evolution of the neuron states and the change in strength of the 
connections (synapses) between neurons. Most of
the prior research on has focused separately on the first dynamical process 
(retrieval
of learned patterns) or the other (learning of patterns) \cite{bo}. 
In this work we deal with the problem of building up  models
where neurons and synapses maintain the same dynamics whether
the network is engaged in computation or in learning new memories; 
the two stages (computation and learning) differ only for
the presence of external stimuli. This problem is important for the description
of the {\it short term} memory in the human brain (see e.g. \cite{br}) and
has been studied mainly in the 
context of Hebbian learning with a decay term \cite{sh}\cite{dh}\cite{gu}:
Shinomoto \cite{sh} presented a rule for the synaptic modification whose stable 
solutions are the Hopfield couplings \cite{ho}; Dong and Hopfield \cite{dh}
considered 
analog neurons in a deterministic   network described by a system of 
differential equations and applied their model to the problem of the 
development of synaptic connections in the visual cortex; 
D'Autilia and Guerra \cite{gu}
studied an ANN with adapting synapses in order to model conditioned reflex 
and rhythm recognition.

We consider here an ANN with asymmetric adapting synapses whose dynamics 
consists of
a Stochastic Learning mechanism \cite{ts}\cite{af}; in particular we refer to 
Amit and Fusi \cite{af} for a discussion on the possible sources 
of stochasticity in the synaptic learning.
We present a dynamical 
theory in terms of deterministic flow equations
for macroscopic order parameters, which holds if the connections are strongly 
diluted. For quenched synapses the 
dynamical approach is exact in the case of strong dilution \cite{de}; 
it may be mentioned, in passing, that recently
Coolen and 
Sherrington described dynamically the fully connected Hopfield model \cite{cs}.
\par
Our model is made of $N$ 
neurons (Ising spins) $s_i (t)\in \{ -1,1\}$, $i=1,\ldots ,N$. For each neuron
$s_i$, 
$K$ input sites $j_1 (i),\ldots ,j_K (i)$ are chosen at random among the $N$ 
sites, and $NK$ synaptic interactions $J_{ij}(t)$ are introduced. We 
assume that the synapses are $n$-states variables 
$J_{ij}\in\{1,{n-3\over{n-1}},\ldots,-1\}$. 
Let us call $\alpha$ the index identifying  the different values of the
synapses, so that $\alpha =1$ implies $J=1$ and $\alpha =n$ implies $J=-1$; 
in general
\begin{equation}
J_{\alpha}={n+1-2\alpha\over {n-1}}\;\;,\;\;\alpha=1,\ldots ,n.
\label{eq:valsyn}
\end{equation}
The dynamical rule for the evolution of the synapses is assumed as follows.
Each synaptic variable $J_{ij}$ connecting $s_i$ with one of its $K$ input
neurons grows by one unit ${2\over n-1}$, i.e. $J_{\alpha}\to J_{\alpha-1}$,
with 
probability $q$ if the product $s_is_j$ of the spins connected by $J_{ij}$ is 
positive; if such product is negative 
the synapse decreases by one unit ($J_{\alpha}\to J_{\alpha +1}$) with the same 
probability. If a synapse is at one of the extreme limits and should be 
pushed off, its value remains unchanged.
\par
A parallel stochastic dynamics with inverse temperature $\beta =T^{-1}$ 
is assumed for 
neurons, where the local field acting on neuron $s_i$ is given by
\begin{equation}
h_i(t)=\sum_j J_{ij}(t) s_j(t) + H_i(t)\; ;
\label{eq:lf}
\end{equation}
the sum is taken over the $K$ input neurons 
and external stimuli are 
represented by local magnetic fields $H_i$. Therefore the rules for the 
parallel dynamics in our model are:
\begin{equation}
s_i(t+1)=sign\left( {1\over 2} +{1\over 2} tanh (\beta h_i(t)) -\eta_i \right)
\label{eq:rule1}
\end{equation}
\begin{equation}
J_{ij}(t+1)=J_{ij}(t)+{2\over{n-1}}s_i(t)s_j(t)\theta (q-\eta_{ij}),
\label{eq:rule2}
\end{equation}
where $\eta_{i}$ and $\eta_{ij}$ are random numbers in $[0,1]$, $\theta$ is the
Heaviside function.
\par
As to the value of $q$, we assume $q=O({1\over K})$. This choice 
deserves a comment. It is known \cite{ts}\cite{af} that if
$q$ is order $\sqrt{{logK\over K}}$ or greater, then it takes only one iteration
to imprint the pattern corresponding to a given neuronic state; since 
the synapses maintain the same dynamics during the retrieval of a previously
learned pattern, such occurrence would destroy the associative capability of 
the network; therefore in the sequel we assume that $q$ is order ${1\over K}$.

Let us first describe qualitatively the behavior of our model in absence of 
external stimuli. At low 
temperature the dynamics drives the system towards one of the $2^N$ 
configurations satisfying $J_{ij}=s_i s_j$ (separable configurations).
If the initial configurations of neurons and synapses are completely
disordered, then at late times the network converges to one state chosen 
among the separable configurations; in other words the network spontaneously 
breaks the symmetry among the separable configurations (a similar spontaneous 
symmetry breaking is discussed in \cite{dh}). If, on the other hand, the 
initial conditions are sufficiently polarized with respect to a given
separable configuration, we expect the network to converge to that 
configuration.

Our aim is to describe macroscopically the evolution of the system by flow 
equations for the macroscopic parameters describing the state of neurons and 
synapses. Let us take $H_i=0$ (no external stimuli on the network) and consider 
an arbitrary pattern $\{\xi_i\}$ for the neurons. We introduce the variables
$\tilde s_i=\xi_i s_i$ and $\tilde J_{ij}=J_{ij}\xi_i\xi_j$, which will be 
useful to measure the 
degree of polarization of neurons and synapses with respect to the pattern 
$\{\xi_i\}$. The statistical ensemble we deal with consists of all the 
hystories of the network with the initial conditions  
$\{s_i(0)\}$ $\{J_{ij}(0)\}$ sampled independently as follows:
\begin{equation}
Prob[s_i(0)=\xi_i]={1\over 2}(1+m_0)
\label{eq:con1}
\end{equation}
\begin{equation}
Prob[J_{ij}(0) =J_\alpha \xi_i\xi_j] =\rho_0 (\alpha )
\label{eq:con2}
\end{equation}
Since the initial conditions are uniform (i.e. independent of the site index 
$i$), the statistical properties of our model remain uniform at every later 
time $t$. We also remark that at $t=0$ both neurons and synapses are 
uncorrelated.
\par
We study the model in the limit $N\to\infty$ with $K$ large and finite.
It can be shown that in such limit the neurons never 
correlate \cite{de}.
Indeed the $K^t$ sites which belong to the
tree of ancestors of a given neuron $s_i$ are all different 
(this is true as long as $t << log N$). Also the correlations 
between synaptic variables can be neglected: the correlation between
two synapses $J_{ij}$ and $J_{ik}$
sharing a common site increases when they are simultaneously updated while  
decreases when one out of the two is updated. The probability for the 
simultaneous updating is order $q^2$, while the probability for the single 
updating is order $q$. Since $q$ is order ${1\over K}$, it follows that, at 
least in the early stage, the synapses can be treated as uncorrelated. In the 
sequel, analyzing in particular the cases $n=2$ and $n=3$, we will show that 
actually the correlation between synapses can ${\it always}$ be neglected.
We also remark 
that $s_j$ is independent of $J_{ij}$ because there are no loops in the tree 
of ancestors of $s_i$. 
Therefore neurons
(as well as synapses) can be treated as independent and identically 
distributed stochastic variables.
\par
We now introduce the order parameter $m(t)=<\tilde s_i(t)>={1\over N}\sum_i
<\tilde s_i(t)>$ for neurons and 
the probability $\rho (\alpha ,t)$  
that $\tilde J_{ij}(t)$ is equal to $J_\alpha$.
We have, from (\ref{eq:con1})(\ref{eq:con2}), 
$\rho (\alpha ,0)=\rho_0 (\alpha)$ and $m(0)=m_0$.
The flow equation for $\rho (\alpha ,t)$ is easily found to be
\begin{equation}
\rho (\alpha ,t+1)=\sum_{\alpha'} T_{\alpha\alpha'}\left( m(t)\right)
\rho (\alpha' ,t),
\label{eq:fl}
\end{equation}
where the $n\times n$ stochastic transition matrix $T$ is tridiagonal with the 
following structure
\begin{equation}
T\left( m(t)\right)=
\left (\begin{array}{ccccc} 
1-a   & b      & 0       &       \ldots        &      0\nonumber\\
a     & 1-q    & b       & \ldots              & 0\\
\ldots&\ldots  &\ldots   &\ldots               &\ldots\\  
0     & \ldots & a       & 1-q                 & b\\
0     &0       & \ldots  & a                   & 1-b\\

\end{array}\right ) \label{matrix}
\end{equation} 
where $a={q\over 2}(1-m^2(t))$ and $b={q\over 2}(1+m^2(t))$.
\par
In order to calculate the flow equation for the order parameter $m(t)$, 
we observe 
that the projection on $\{\xi\}$ of local field (\ref{eq:lf}) can be written as
\begin{equation}
\tilde h_i(t)=\xi_i h_i(t)=\sum_{j} \tilde J_{ij}(t)\tilde s_j(t)
=\sum_j 
J'_{ij}(t),
\label{eq:lf1}
\end{equation}
where the variable $J'_{ij}=\tilde J_{ij} \tilde s_j$ is the contribution to 
the synaptic input from a single input neuron.
The variable $\tilde h$ (the total synaptic input) 
is the sum of $K$ independent and identically distributed stochastic variables 
$J'$. The 
probability distribution for $J'$ reads:
\begin{equation}
\rho'(\alpha ,t)= Prob \; 
[J'_{ij}(t)=J_\alpha] ={1\over 2}(1+m(t))\rho(\alpha,t)+
{1\over 2}(1-m(t))\rho(n+1-\alpha,t),
\label{eq:distr}
\end{equation}
where the first term on the RHS of eq.(\ref{eq:distr}) is the probability that
$\tilde s=1$ and $\tilde J =J_{\alpha}$, the second term is the probability 
that $\tilde s=-1$ and $\tilde J =-J_{\alpha}$.
The stochastic variable $\tilde h_i(t)$ can assume $(n-1)K+1$ possible discrete
values in [$-K$,$K$]; its distribution ${\cal H}_t (a)=
Prob\; [\tilde h_i(t) =a]$
can be calculated by performing $K$ times the convolution of the distribution
(\ref{eq:distr}).
For example, in the case $n=2$, we have
\begin{equation}
{\cal H}_t(a)= {K\choose {K-a\over 2}} {\rho'(1,t)}^{K+a\over 2} 
{\rho'(-1,t)}^{K-a\over 2}.
\label{eq:distr1}
\end{equation}
It is now straightforward to evaluate, in the general case, 
the flow equation for $m(t)$ using eqs.(\ref{eq:rule1}):
\begin{equation}
m(t+1)=\sum_{a=-K}^K {\cal H}_t(a)\;  tanh(\beta a) =\; 
<tanh(\beta\tilde h)>_{{\cal H}_t}
\label{eq:fl2}
\end{equation}
Flow equations (\ref{eq:fl}) and (\ref{eq:fl2}), with initial conditions
$m_0$ and $\rho_0(\alpha )$, describe the coupled dynamics of $m(t)$ 
and $\rho(\alpha ,t)$. 
\par
After this general introduction and the description of the main dynamical 
equations, we will devote the next section to study the fixed 
points of the dynamics for our model. In section 3 we will report the results 
obtained simulating numerically the flow equations in the cases 
$n=2$ and $n=3$. In section 4 it is shown that the correlations among synapses 
can always be neglected. In section 5 we study the 
learning properties of the network. In section 6 the 
simulations of an ANN with adapting synapses are studied and compared to the 
theory. Section 7 summarizes the conclusions.

\vskip 2 truecm
\par\noindent
{\bf 2. Stationary solutions.}
\par
Let us now  discuss the stationary solutions of the flow equations. 
First of all we 
observe that the invariant distribution with respect to (\ref{eq:fl}) is given 
by
\begin{equation}
\rho_m(\alpha)={2m^2\over {{(1+m^2)}^n -{(1-m^2)}^n}}
{(1-m^2)}^{\alpha -1} {(1+m^2)}^{n-\alpha}
\label{eq:stat}
\end{equation}
We remark that the stationary 
distribution does not depend on $q$. We call ${\cal H}^{(m)} (a)$ the 
probability 
distribution of $\tilde h$ corresponding to $\rho_m$. For stationary solutions 
the following equation holds:
\begin{equation}
m=<tanh(\beta\tilde h)>_{{\cal H}^{(m)}}
\label{eq:stat2}
\end{equation}
It can be shown that $m=0$ is always a stable solution for (\ref{eq:stat2}), 
and it is unique for small values of $\beta$ (high temperature).
As $\beta$ increases, equation (\ref{eq:stat2}) displays a first-order 
transition, i.e. for $\beta > \beta_c (K)$ two solutions with $m>0$ appear 
discontinuously, the one with larger $m$ being locally stable for variations 
in $m$. This 
solution corresponds to a state which deviates only slightly from pattern 
$\{\xi\}$; $m=1$ is a stationary and stable solution only in the limit 
$\beta\to\infty$. We have thus shown that the model possesses a stable fixed 
point
in correspondence to any pattern $\{\xi\}$. If the initial conditions $m_0$ 
and $\rho_0(\alpha )$ are sufficiently polarized with respect to $\{\xi\}$, 
then, by iterating the flow equations,
the stable solution $m_s>0$ is asymptotically achieved, i.e.:
\begin{equation}
\lim_{t\to\infty} m(t)=m_s\;\;\;\;\;\;\;\;\;\lim_{t\to\infty} \rho(\alpha ,t)
=\rho_{m_s}(\alpha)
\label{eq:stat3}
\end{equation}
In the case of a large number of connections (large $K$) a useful approximation
can be used.
Since the local field $\tilde h_i(t)$ is the sum of $K$ independent and 
identically distributed stochastic 
variables, we use the central limit theorem to approximate it by a Gaussian 
random variable with mean  and variance given respectively by
\begin{equation}
\mu(t)=Km(t)<\tilde J(t)>
\label{eq:limc33}
\end{equation}
and
\begin{equation}
\sigma^2(t)=K\left(<\tilde J^2(t)>
-<\tilde J(t)>^2 m^2(t)\right)
\label{eq:limc}
\end{equation}
The flow equation (\ref{eq:fl2}) can be approximated as follows:
\begin{equation}
m(t+1)=\int {dz\over \sqrt{2\pi}} e^{-{1\over 2}z^2} tanh\big[ \beta\left(\mu(t)
+\sigma(t) z\right)\big].
\label{eq:fl3}
\end{equation}
In the zero temperature limit ($\beta\to\infty$) equation (\ref{eq:fl3}) reads
\begin{equation}
m(t+1)= erf\;\;\left({\mu(t)\over \sigma(t)}\right),
\label{eq:fl4}
\end{equation}
where ${\mu\over \sigma}$ may be seen as the signal-to-noise ratio 
(SNR) for the synaptic input of a neuron.

\vskip 2 truecm
\par\noindent
{\bf 3. Analysis of the cases $n=2$ and $n=3$.} 
\par
In this section we study numerically the behavior of the flow equations 
(\ref{eq:fl}) and (\ref{eq:fl3}) in the cases $n=2$ and $n=3$. 
The probability distribution for two-state
synapses is determined by its average ${\cal J}(t)=<\tilde J(t)>$ and flow 
equations reduce to:
\begin{equation}
{\cal J}(t+1)=(1-q){\cal J}(t)+qm^2(t)
\label{eq:fl5}
\end{equation}
\begin{equation}
m(t+1)=\int {dz\over \sqrt{2\pi}} e^{-{1\over 2}z^2} tanh\big[\beta\left(K
{\cal J}(t)m(t)+\sqrt{K\left(1-{\cal J}^2(t)m^2(t)\right)}\; z\right)\big].
\label{eq:abb}
\end{equation}
The stationarity conditions are
\begin{equation}
{\cal J}=m^2
\label{eq:fl6}
\end{equation}
\begin{equation}
m=\int {dz\over \sqrt{2\pi}} e^{-{1\over 2}z^2} tanh\;\;\beta\left(Km^3
+\sqrt{K(1-m^6)}\; z\right)
\label{eq:flm}
\end{equation}
Equation (\ref{eq:flm}) 
displays a first-order transition for $\beta$ greater 
than a critical coupling which has the behavior $\beta_c(K)\sim {2.017\over 
K}$ for large $K$. Hence for $\beta >\beta_c$ equations (\ref{eq:fl5})
(\ref{eq:abb}) have two 
stable fixed points, a fixed point with $m>0$ and the trivial one. By 
numerical simulations we found that the recurrence equations (\ref{eq:fl5})
(\ref{eq:abb}) never 
show complex behavior and by iterating them, starting from any initial 
condition, it always happens that one of the 
two stable fixed points is approached.
\par
It is interesting to compare the retrieval capability of our neural network
with adapting synapses to the case of a strongly diluted network with 
fixed synapses, where each neuron has $K$ random input sites, described by 
equation (\ref{eq:abb}) with ${\cal J}(t)={\cal J}_0$ for every $t$.
The process of retrieval 
in our model will be referred to as {\it adaptative retrieval} (AR)
whereas the 
retrieval with fixed synapses  will be called {\it fixed synapses retrieval} 
(FSR). We remark that FSR actually corresponds to the early stage of AR, due to 
the very low value of $q$.
In fig.1 we depicted the asymptotic value $m(\infty )$ of the order 
parameter versus the initial condition ${\cal J}_0$ for the two cases and 
various 
temperatures. We see that in the case of AR a first order transition occurs;
in other words, for ${\cal J}_0$ greater than a threshold value 
${\cal J}_{th}$, the non-trivial fixed point of equation (\ref{eq:flm}) is 
asymptotically achieved; on the other hand
a second order transition occurs in the case of FSR. For ${\cal J}_0$ 
greater than ${\cal J}_{th}$ and ${\cal J}_0-{\cal J}_{th}$ small, AR performs 
better than FSR. We observe that a threshold value exists also for $m_0$,
i.e. for $m_0$ smaller than a threshold value $m_{th}$ 
(numerically very small), the flow 
equations (\ref{eq:fl5})(\ref{eq:abb}) lead to the trivial fixed point even for 
${\cal J}_0=1$.
\par
Let us now  consider the case $n=3$, when synapses take values in 
$\{1,0,-1\}$. Two independent parameters are needed in order to specify the 
distribution of the synapses; let us define
$\rho^{+}(t)=\rho (1,t)$, $\rho^{-}(t)=\rho (-1,t)$ and 
$\rho^{0}(t)=\rho (0,t)$. The stationarity conditions can be computed in this 
case with the results:
\begin{equation}
\rho^{+}={{(1+m^2)}^2\over {3+m^4}}\;\;\;\;\;\;\;\;\rho^{0}={1-m^4\over {3+m^4}}
\;\;\;\;\;\;\;\;\rho^{-}={{(1-m^2)}^2\over {3+m^4}}
\label{eq:stas}
\end{equation}
\begin{equation}
m=\int {dz\over \sqrt{2\pi}} e^{-{1\over 2}z^2} tanh\big[ \beta\left(Km(
\rho^{+}-\rho^{-})
+\sqrt{K(1-\rho^{0}-m^2{(\rho^{+}-\rho^{-})}^2)}\; z\right)\big].
\label{eq:star}
\end{equation}
The same qualitative description of the $n=2$ case applies to the behavior 
of the flow equations in this case. The critical coupling has the behavior
$\beta_c (K)\sim {1.8\over K}$ for large $K$.
In fig.2 we depict, for three values of the temperature, the threshold value
of $\rho_D =\rho^{+}-\rho^{-}$ versus $\rho^{0}$. The threshold $\rho_D^{th}$
is such that the non-trivial fixed point is achieved if the initial conditions 
satisfy $\rho_D > \rho_D^{th}$. 
\vskip 2 truecm
\par\noindent
{\bf 4. The synaptic correlations.}
\par
In this section we show that, during the time evolution of the 
network, the correlations among the synapses can be neglected. This assumption 
is tested by taking into account the correlation
between pairs of synapses sharing a common site. Consider the local field 
$\tilde h_i$
acting on the neuron $s_i$ given by equation (\ref{eq:lf1}). The mean and 
variance of $\tilde h_i$ are given by:
\begin{equation}
\mu=K<\tilde J> m
\label{eq:cor1}
\end{equation}
\begin{equation}
\sigma^2 =K\left(<\tilde J^2> - <\tilde J>^2 m^2\right)+K(K-1)m^2{\cal C}
\label{eq:cor2}
\end{equation}
where ${\cal C}=<\tilde J_{ij}\tilde J_{ik}> - <\tilde J_{ij}><\tilde J_{ik}>$
is the pair synaptic correlation. The first term on the RHS of 
eq.(\ref{eq:cor2}) is the variance of the local field which is computed as if 
the synapses were uncorrelated.
\par
In the $n=2$ case we consider the probability distribution for pairs of 
synapses $\{\tilde J_{ij},\tilde J_{ik}\}$ sharing a common site, which evolves 
according to a $4\times 4$ transition matrix. After a little algebra we obtain 
the flow equations for ${\cal J}$ (=$<\tilde J_{ij}>=<\tilde J_{ik}>$) and
${\cal C}$:
\begin{equation}
{\cal J}(t+1)=(1-q){\cal J}(t)+qm^2(t)
\label{eq:cor3}
\end{equation}
\begin{equation}
{\cal C}(t+1)=(1-q)^2{\cal C}(t)+q^2\left(m^2(t)-m^4(t)\right)
\label{eq:cor4}
\end{equation}
while equation (\ref{eq:fl3}) can be used for the evolution of the order 
parameter $m(t)$; the initial condition for the correlation is ${\cal C}(0)=0$.
Equation (\ref{eq:cor4}) has a clear meaning: the pair correlation does not 
change if the two synapses are not updated (probability $(1-q)^2$); it goes to 
zero if one out of the two is updated (probability $2q(1-q)$); it 
becomes $m^2-m^4$ if both synapses are updated (probability $q^2$).
We studied numerically eqs. (\ref{eq:cor3}), (\ref{eq:cor4}) and (\ref{eq:fl3})
and found that the second term on the RHS of equation (\ref{eq:cor2}) is always
very small compared to the first term. In fig.3 we depict the time 
evolution of the order parameter, for a particular set of initial conditions,
in two cases: taking into account the pair correlations and not. 
The two curves are
indistinguishable; the same behavior is observed for any choice of the initial 
conditions. In fig.3 we also depict the time evolution of the two contributions 
to the variance of the local field, given by eq.(\ref{eq:cor2}). 
The contribution 
due to the synaptic correlations is always very small and reaches a maximum 
value during the retrieval process. The stationarity conditions for eqs.
(\ref{eq:cor3}) and (\ref{eq:cor4}) read:
\begin{equation}
{\cal J}=m^2,\;\;\;\;\;\;{\cal C}={q\over 2-q}(m^2-m^4),
\label{eq:statt}
\end{equation}
and, at leading order in $K$, the equation for the fixed points is:
\begin{equation}
m=\int {dz\over \sqrt{2\pi}} e^{-{1\over 2}z^2} tanh\;\;\beta\left(Km^3
+\sqrt{K(1-m^6+{m^4-m^6\over 2})}\; z\right)
\label{eq:stott}
\end{equation}
to be compared with the equation which is computed as if the synapses
were uncorrelated, i.e. eq.(\ref{eq:flm}). We see that the distance between the
non-trivial fixed points of eqs.(\ref{eq:stott}) and (\ref{eq:flm}) 
vanishes for large $K$. Hence the synaptic correlations can be neglected also
with respect to the equilibrium properties of the network.
\par
We tested the assumption also in the $n=3$ case (see the Appendix).
We find that also  in this case the 
contribution to $\sigma^2$ due to the correlations is 
always very small and can be neglected. In fig.4 we depicted the time evolution 
of the order parameter $m$, in a particular history, calculated in two cases,
i.e. with and without correlations. 
Also in this case the two curves are indistinguishable.
\vskip 2 truecm
\par\noindent

{\bf 5. The learning stage.}
\par
The learning stage is characterized by the presence of external stimuli 
represented by local magnetic fields in equation (\ref{eq:lf}). 
We recall the results obtained in \cite{af} about the capacity of a stochastic
learning network with  $q$ of order $\sqrt{{logK\over K}}$ and with $n$ 
arbitrary. 
Let us suppose that 
the network receives an uninterrupted flow of uncorrelated patterns to be 
learned:
\begin{center}
$\ldots\;,\;\{\xi\}_{-2}\;,\;\{\xi\}_{-1}\;,\;\{\xi\}_{0}\;,\;\{\xi\}_{1}\;
,\;\{\xi\}_{2}\;,\;\ldots$
\end{center}
In \cite{af} it is shown, by a signal-to-noise analysis, 
that the network acts as a palympsest 
\cite{na},\cite{pa}, i.e. patterns learned far in the past are erased by new 
patterns, and that the maximum number of patterns which can be stored is 
proportional to $\sqrt{K}$ \cite{preci}. 
\par 
In our case $q$ is order ${1\over K}$ and each 
pattern $\{\xi\}_p$ ($p=0,\pm 1,\pm 2,\ldots$)
has to be presented at least for $\ell =O\left(\sqrt{K\;logK}\right)$ 
iterations in order to be stored. 
Let us consider for example the pattern $\{\xi\}_0$, and let us set $t=0$ in 
correspondence of the first presentation of $\{\xi\}_0$.
Initially the synapses have 
a completely random distribution with respect to $\{\xi\}_0$, i.e.
$\rho (\alpha,0)={1\over n}$. During the presentation of pattern $\{\xi\}_0$
the synapses are 
polarized due to the action of matrix (\ref{matrix}) with $m=1$, because 
$\{s\}=\{\xi\}_0$.
Subsequently the network begins to learn other patterns and the synaptic 
distribution becomes depolarized by the action of matrix (\ref{matrix})
with $m=0$. After the presentation of $z$ other patterns the synaptic 
distribution is given by
\begin{equation}
\rho \left( t=(z+1)\ell\right)=T^{z\ell}_q (m=0)T^{\ell}_q (m=1) \rho (0)
\label{eq:lear}
\end{equation}
where the dependence of $T$ on $q$ has been made explicit.
Incidentally we remark that during the learning stage the correlation among 
synapses is identically zero due to the $\pm$ symmetry. For fixed $m$ and
$q$ order ${1\over K}$ we have
\begin{equation}
T^{\ell}_q = T_{\ell q} +O(\ell^2 q^2) = T_{\ell q} +
O\left({log K\over K}\right) 
\label{eq:lear1}
\end{equation}
as can be easily checked performing the Taylor expansion of the matrix $T_q$
in a neighborhood of $q=0$.
Therefore rescaling the time by the factor $\ell$ and changing $q$ into
$\tilde q=\ell q$ leads to the same problem studied in \cite{af}.
Hence we conclude that our network is a palympsest and its capacity
is proportional to $\sqrt{K}$. 
\vskip 2 truecm
\par\noindent
{\bf 6. Simulations.}
\par
Our theory works as long as $t$ is small with respect to $log N$, so as to
avoid the effects due to the neuronic correlations. In order to build up 
finite-size networks, where the neuronic correlations can be neglected 
for longer times, and to test our dynamical equations, we adopt 
the following approach: we implement a model of $N$ neurons, where each neuron
has $M$ input neurons chosen at random ($M<<N$). The $NM$ synapses obey the 
stochastic learning rules and, for every neuron $s_i$ and for every time $t$, 
$K$ input neurons are chosen at random among the $M$ corresponding to $s_i$
and only those $K$ inputs contribute to the local field acting on $s_i$
at time $t$. This is a generalization of our model (recovered
when $K=M$) which is ruled by the same dynamical equations.
We implemented a $N=10,000$ network with $M=200$, $K=21$ and two-states 
synapses.
The time evolution of the macroscopic parameters of the system
is found to be in agreement with the theoretical estimate until $t\sim 20$.
In fig.5 we depict the time evolution of the order parameter and of the 
synaptic distribution and compare it with the theoretical estimate; the initial 
conditions were $m_0=1$ and ${\cal J}_0=0.3$. The agreement with the theory has 
been tested with many different initial conditions. 
We also implemented a network with $N=10,000$ and $M=K=21$ but the time 
evolution in this case was in agreement with the theory only for a few time
steps.

\vskip 2 truecm
\par\noindent
{\bf 7. Conclusions.}
\par
In this paper we have considered a neural network with clipped synapses 
and stochastic
learning rules whose learning capabilities have been studied in \cite{af}.
We extend the analysis of this model and investigate the consequences of the 
synaptic dynamics in the process of retrieval. We find that, in order to 
preserve the associative capability of the system, the synaptic transition 
probability $q$ must be very small; moreover, for strong dilution, 
the dynamics of 
the network can be analitically calculated because for very small values of $q$
the correlations among the synaptic variables can be neglected.
As to the learning properties, the network acts as a palympsest and the maximum 
number of storable patterns coincides with the result obtained in \cite{af},
the only difference being that a pattern has to be presented for many 
iterations in order to be stored.
\par
In this framework the two stages (computation and learning) differ for
the duration of the external stimulus corresponding to a given pattern. If
the pattern is presented for a sufficiently long time, the network stores it 
in the synaptic couplings. If the pattern (or a damaged version of it)
is presented for few iterations, the 
dynamics of the network is capable to reconstruct the pattern provided it 
had been learned previously (unless new external stimuli impinge on the network
during the retrieval).
\par
It would be interesting to study the coupled dynamics of neurons and synapses,
assuming a stochastic learning mechanism, in a fully connected network.
\vskip 2 truecm
\par\noindent
{\bf Appendix}
\par
We write here the flow equations, in the $n=3$ case, which take into account
the pair synaptic correlations. 
Let us consider the distribution 
of a pair of synapses $\{\tilde J_{ij},\tilde J_{ik}\}$ sharing a common site,
which evolves according to a $9\times 9$ transition matrix.
We define
$$\rho^{+}=Pr[\tilde J_{ij}=1]=Pr[\tilde J_{ik}=1]$$
$$\rho^{-}=Pr[\tilde J_{ij}=-1]=Pr[\tilde J_{ik}=-1]$$
$$\rho^{0}=Pr[\tilde J_{ij}=0]=Pr[\tilde J_{ik}=0]$$ 
$$\rho^{+0}=Pr[\tilde J_{ij}=1,\tilde J_{ik}=0]=
Pr[\tilde J_{ij}=0 ,\tilde J_{ik}=1]$$
$$\rho^{+-}=Pr[\tilde J_{ij}=1 , \tilde J_{ik}=-1]=
Pr[\tilde J_{ij}=-1 , \tilde J_{ik}=1]$$
$$\rho^{-0}=Pr[\tilde J_{ij}=-1 , \tilde J_{ik}=0]=
Pr[\tilde J_{ij}=0 , \tilde J_{ik}=-1].$$
Next, we have 
$$<\tilde J^2>=\rho_{+}+\rho_{-}$$ 
$$<\tilde J>=\rho_{+}-\rho_{-}$$
$${\cal C}=<\tilde J_{ij}\tilde J_{ik}> - <\tilde J_{ij}><\tilde J_{ik}>=
\rho^{+}+\rho^{-}-\rho^{+0}-\rho^{-0}-4\rho^{+-}-
(\rho^{+}-\rho^{-})^2.$$
The flow equations for $\rho^{+},\rho^{-},\rho^{0}$ are given by (\ref{eq:fl}),
while the flow equation for $m(t)$ is (\ref{eq:fl3}). The other flow equations
are:
\begin{equation}
\rho^{+0}(t+1)=(a_1 -a_3)\rho^{+0}(t)+(a_2-a_3+a_4-a_5)\rho^{+-}(t)+(a_4-a_2)
\rho^{-0}(t)+(a_3+a_5)\rho^{+}(t)+a_2\rho^{0}(t)
\label{eq:cor5}
\end{equation}
\begin{equation}
\rho^{+-}(t+1)=a_3\rho^{+0}(t)+(a_1+a_2+a_3+a_5)\rho^{+-}(t)+a_2\rho^{-0}(t)+
a_5\rho^{0}(t)
\label{eq:cor6}
\end{equation}
\begin{equation}
\rho^{-0}(t+1)=(a_5-a_3)\rho^{+0}(t)+(a_3-a_2)\rho^{+-}(t)+(a_1-a_2)\rho^{-0}
(t)+a_3\rho^{0}(t)+(a_2+a_5)\rho^{-}(t)
\label{eq:cor7}
\end{equation}
where $a_1=(1-q)^2$, $a_2={1\over 2}q(1-q)(1+m^2)$, 
$a_3={1\over 2}q(1-q)(1-m^2)$, $a_4={1\over 4}q^2(1+3m^2)$ and $a_5={1\over 4}
q^2(1-m^2)$.

\newpage
\Large
\par\noindent
{\bf Figure Captions}
\normalsize
\vskip 1 truecm
\par\noindent
Figure 1: Numerical evaluation of the asymptotic order parameter, versus the 
initial polarization ${\cal J}_0$ of the synapses, with  
$K=100$, $q=0.01$, $n=2$ and three different values of the temperature.
The continuous line and the dotted one correspond respectively to
AR and FSR (see the definitions in the text).
\vskip 0.5 truecm
\par\noindent
Figure 2: The threshold value of $\rho_D$ is plotted versus $\rho^{0}$ in the 
case $n=3$, $K=100$, $q=0.01$ and with three different values of the
temperature.
\vskip 0.5 truecm
\par\noindent
Figure 3: The time evolution of the order parameter evaluated by the flow 
equations which neglect the correlations ($\bullet$) and by the flow equations
which take into account the pair synaptic correlation ($\circ$) in the case 
$n=2$: the two curves are indistinguishable. 
The initial 
conditions are $m_0=0.88$, ${\cal J}_0=0.55$; $K=100$, $q=0.01$, $\beta=0.03$.
In the figure we also depict the time evolution of the two contributions to the 
variance of the local field (see the text): the contribution due to the 
synaptic correlations, i.e. $(K-1)m^2{\cal C}$ ($\diamondsuit$),
compared to the term which is computed as if the synapses were uncorrelated
, i.e. $1-{\cal J}^2 m^2$ ($\triangle$). 
\vskip 0.5 truecm
\par\noindent
Figure 4: The time evolution of the order parameter evaluated by the flow 
equations which neglect the correlations ($\bullet$) and by the flow equations
which take into account the pair synaptic correlation ($\circ$) in the case 
$n=3$: as in Fig.3 the two curves are indistinguishable. 
The initial 
conditions are $m_0=0.8$, $\rho_D=0.3$, $\rho^{0}=0.33$; $K=100$, $q=0.01$, 
$\beta=0.05$.
In the figure we also depict the time evolution of the two contributions to the 
variance of the local field (see the text): the contribution due to the 
synaptic correlations, i.e. $(K-1)m^2{\cal C}$ ($\diamondsuit$), 
compared to the term computed as if the synapses were uncorrelated
, i.e. $1-\rho^{0}-{(\rho^{+}-\rho^{-})}^2 m^2$ ($\triangle$). 
\vskip 0.5 truecm
\par\noindent
Figure 5: Time evolution of the order parameter $<\tilde s>$ and the mean of 
the synapses $<\tilde J>$ in a ANN with two-states adapting synapses. 
The numerical results for $<\tilde s>$ ($\bullet$) and $<\tilde J>$ 
($\diamondsuit$) are averaged over $80$
hystories of a network with $N=10,000$, $M=200$, $K=21$, $q=0.01$
and zero temperature; the 
initial conditions are $m_0=1$ and ${\cal J}_0=0.3$. The theoretical estimate
by the flow equations is represented by $\circ$ ($<\tilde s>$) and by 
$\triangle$ ($<\tilde J>$). 


\begin{thebibliography}{99}
\bibitem{preci} Actually in ref.[8] a fully connected network is considered
and the capacity is proportional to $\sqrt{N}$, where $N$ is the number of 
neurons in the network.
The same analysis holds in a diluted network where each neuron is connected 
to $K$ random input neurons and the capacity in this case is proportional to 
$\sqrt{K}$.
\end{thebibliography}
\end{document}